\documentstyle[12pt]{article}
\newcommand{\beq}{\begin{equation}}
\newcommand{\eeq}{\end{equation}}

\newcommand{\beqn}{\begin{eqnarray}}
\newcommand{\eeqn}{\end{eqnarray}}

\newcommand{\ba}{\bar \alpha}
\newcommand{\bb}{\bar \beta}

\begin{document}

\begin{center}{\Large \bf Proton polarizability contribution
to hydrogen Lamb shift}\\
\vspace{1.0cm}      

{\bf I.B. Khriplovich}  \\
Budker Institute of Nuclear Physics, 630090 Novosibirsk, Russia\\
\vspace{0.5cm}
{\bf R.A. Sen'kov}\\
Novosibirsk University, 630090 Novosibirsk, Russia\\
                         
\vspace{1.0cm}
\end{center}

\begin{abstract}                                    
The correction to the hydrogen Lamb shift due to the proton electric
and magnetic polarizabilities is expressed analytically through their
static values, which are known from experiment. The numerical value
of the correction is $\;-\;71 \pm \,11 \pm\,7$ Hz.
\end{abstract}

\bigskip

High experimental precision attained in the hydrogen and deuterium
spectroscopy (see, e.g., \cite{han}) stimulates considerable
theoretical activity in this field. In particular, the deuteron
polarizability contribution to the Lamb shift in deuterium was
calculated in Refs. [2 -- 9]. An estimate of the proton
polarizability contribution to the Lamb shift in hydrogen was made in
Ref. \cite{pach}. A special feature of the corrections obtained in
Refs. [2 -- 9] is that they contain logarithm of the ratio of a
typical nuclear excitation energy to the electron mass, $\ln \bar
E/m_e$. In fact, the contribution of the nuclear electric
polarizability to the Lamb shift had been obtained with the
logarithmic accuracy in Ref. \cite{eric} for an arbitrary nucleus.

In the present note we consider the problem of the proton
polarizability correction to the Lamb shift in hydrogen. The typical
excitation energy for the proton $\bar E_p \sim 300$ MeV is large as
compared to other nuclei (to say nothing of the deuteron). So, the
logarithm $\ln \bar E/m_e$ is not just a mere theoretical parameter,
it is truly large, about $6 - 7$, which makes the logarithmic
approximation quite meaningful quantitatively.  As distinct from Ref.
\cite{eric}, we take into account in our final formula not only the
electric polarizability $\bar \alpha$, but as well the magnetic one
$\bar \beta$ (though it does not very much influence the result
numerically).

In our calculation we follow closely the approach of Ref. \cite{khr}.
In particular, we use the gauge $A_0=0$ for virtual photons, so that
the only nonvanishing components of the photon propagator are
$ D_{im}=d_{im}/k^2,\;\; d_{im}=\delta_{im}- k_i k_m/{\omega}^2\;\;
(i,m=1,2,3)$. The electron-proton forward scattering amplitude, we are
interested in, is 
\begin{equation}\label{ep}
T=4\pi i \alpha \int \frac{d^4k}{(2\pi)^4}\;D_{im}D_{jn}
       \frac{\gamma_i({\hat l}-{\hat k}+m_e)\gamma_j}{k^2-2lk}\;M_{mn}.
\end{equation}
Here $l_{\mu}=(m_e,0,0,0)$ is the electron momentum. The nuclear-spin
independent Compton forward scattering amplitude, which is of
interest to as, can be written as
\begin{equation}\label{gp}
 M=\ba(\omega^2,{\bf k^2}){\bf E^*E}+\bb(\omega^2,{\bf k^2}){\bf B^*B}
 \;=M_{mn}e_me_n{}^*,
\end{equation}
where $\ba$ and $\bb$ are the nuclear electric and magnetic
polarizabilities, respectively.  The structure $ \gamma_i({\hat
l}-{\hat k}+m_e)\gamma_j$ in (\ref{ep}) reduces to $ -
\omega\delta_{ij} $.  Perhaps, the most convenient succession of
integrating expression (\ref{ep}) is as follows: the Wick
rotation; transforming the integral over the Euclidean $\omega$ to
the interval $(0,\,\infty)$; the substitution ${\bf k}\rightarrow
{\bf k}\,\omega$. Then the integration over $\omega$ is easily
performed with the logarithmic accuracy:
\begin{eqnarray}\label{c}
{\int_0}^{\infty}\frac{d\omega^2}{\omega^2+4{m_e}^2/(1+{\bf k}^2)^2}
  \;\left[(3+2{\bf k}^2+{\bf k}^4)\ba(-\omega^2,-\omega^2{\bf k}^2)-
          2{\bf k}^2\bb(-\omega^2,-\omega^2{\bf k}^2)
    \right]\;\\
    =\;\left[(3+2{\bf k}^2+{\bf k}^4)\ba(0)
                   -2{\bf k}^2\bb(0)\right]\;
    \ln\frac{\bar E ^2}{m_e^2}.\nonumber
\end{eqnarray}
The crucial point is that within the logarithmic approximation both
polarizabilities $\ba$ and $\bb$ in the lhs can be taken at
$\omega=0$, ${\bf k}^2=0$. The final integration over $d^3{\bf k}$
is trivial.

0The resulting effective operator of the electron-proton interaction
(equal to $-T$) can be written in the coordinate representation as
\begin{equation}\label{d}
V=-\,\alpha m_e\, [\,5\ba(0)-\bb(0)\,]\;\ln\frac{\bar E}{m_e}\;\delta({\bf r}).
\end{equation}
This expression applies within the logarithmic accuracy for arbitrary
nuclei. As it was mentioned above, it differs from the result
obtained earlier in Ref. \cite{eric} by the account for the magnetic
polarizability $-\bb(0)$ only.

As refers to hydrogen, the experimental data on the proton
polarizabilities, which follow from the Compton scattering, can be
summarized as follows \cite{mac}:
\[ \ba_p(0)+\bb_p(0)=(14.2 \pm 0.5)\times 10^{-4}\; \mbox{fm}^3; \]
\beq\label{co}
\ba_p(0)-\bb_p(0)=(10.0 \pm 1.8)\times 10^{-4}\; \mbox{fm}^3.
\eeq 
Now, 
\beq\label{co1}
5\,\ba_p(0)-\bb_p(0)\,=\,2\,[\,\ba_p(0)+\bb_p(0)\,]\,
+\,3\,[\,\ba_p(0)-\bb_p(0)\,]
=(58.4 \pm 5.3)\times 10^{-4}\; \mbox{fm}^3.
\eeq 
The errors are added in quadratures.

Finally, at $\bar E_p \sim 300$ MeV the proton polarizability correction 
to the hydrogen $1S$ state is
\beq\label{res}
-\,71\,\pm\,11\,\pm\,7\; \mbox{Hz}.
\eeq
Here the first error is that of the logarithmic approximation, which
we estimate as 15\%. The second one originates from the values of the 
polarizabilities.

The corresponding estimate presented in Ref. \cite{pa2} differs from
our result by the factor at $\ba_p(0)$ (2 instead of 5) and by the
absence of $\bb_p(0)$.

The obtained correction (\ref{res}) is not so far away from the
accuracy expected soon in the measurements of the isotope shift
between deuterium and hydrogen $1S - 2S$ transitions.

\bigskip

We are grateful to A.I. Milstein and A.A. Pomeransky for useful
discussions. We wish to thank also J.L. Friar who attracted our
attention to Ref. \cite{eric}. I.Kh. acknowledges the support by the
Russian Foundation for Basic Research through grant No. 95-02-04436-a.


\bigskip

\begin{thebibliography}{99}
\bibitem{han}K. Pachucki, D. Leibfried, M. Weitz, A. Huber, W. K\"onig,
              and T.W. H\"ansch, J.Phys. B {\bf 29}, 177 (1993).
\bibitem{pach}K. Pachucki, D. Leibfried, and T.W. H\"ansch, Phys. Rev.
              A {\bf 48}, R1 (1993).
\bibitem{pa2} K. Pachucki, M. Weitz, and T.W. H\"ansch, Phys. Rev.
              A {\bf 49}, 2255 (1994).
\bibitem{li} Yang Li and R. Rosenfelder, Phys.Lett. B {\bf 319}, 7 (1993);
             {\it ibid}, {\bf 333}, 564 (1994).
\bibitem{lei} W. Leidemann, and R. Rosenfelder, Phys. Rev.
              C {\bf 51}, 427 (1995).
\bibitem{mar} J. Martorell, D.W.L. Sprung, and D.C. Zheng, Phys. Rev.
              C {\bf 51}, 1127 (1995).
\bibitem{khr} A.I. Milstein, S.S. Petrosyan, and I.B. Khriplovich,
              Zh.Eksp.Teor.Fiz. {\bf 109}, 1146, 1996 
              [Sov.Phys. JETP {\bf 82}, 616 (1996)].
\bibitem{fri} J.L. Friar, and G.L. Payne, nucl-th/9702019; Phys. Rev.
              C, in press (1997).
\bibitem{fri1} J.L. Friar, and G.L. Payne, nucl-th/9704032.
\bibitem{eric}J. Bernab\'eu and T.E.O. Ericson, Z.Phys. A {\bf 309}, 
              213 (1983).
\bibitem{mac} B.E. MacGibbon, G. Garino, M.A. Lucas, A.M. Nathan, 
              G. Feldman, and B. Dolbikin, Phys. Rev. C {\bf 52}, 
              2097 (1995).

\end{thebibliography}
\end{document}